\documentstyle[12pt,aasms4]{article}

\def\arcsecpoint{$''\!.$}
\received{}
\accepted{}

\slugcomment{submitted to {\it The Astrophysical Journal}}

\lefthead{Kraemer, Crenshaw, George, Netzer, Turner, \& Gabel}
\righthead{Variable UV Absorption in the 
Seyfert 1 Galaxy NGC 3516: The Case for Associated UV and X-ray
Absorption}

\begin{document}

\title{Variable UV Absorption in the 
Seyfert 1 Galaxy NGC 3516: The Case for Associated UV and X-ray 
Absorption\altaffilmark{1}}

\author{S. B. Kraemer\altaffilmark{2},
D. M. Crenshaw\altaffilmark{3},
I.M. George\altaffilmark{4,5},
H. Netzer\altaffilmark{6},
T.J. Turner\altaffilmark{4,5},
\& J.R. Gabel\altaffilmark{2}}

\altaffiltext{1}{Based on observations made with the NASA/ESA Hubble Space
Telescope. STScI is operated by the Association of Universities for Research in
Astronomy, Inc. under the NASA contract NAS5-26555. }
  
\altaffiltext{2}{Catholic University of America,
and Laboratory for Astronomy and Solar Physics, NASA's Goddard Space Flight 
Center, Code 681,
Greenbelt, MD  20771; stiskraemer@yancey.gsfc.nasa.gov.}

\altaffiltext{3}{Department of Physics and Astronomy, Georgia State University,
Atlanta, GA 30303.}

\altaffiltext{4}{Joint Center for Astrophysics, Physics Dept., University
of Maryland Baltimore County, 1000 Hilltop Circle, Baltimore, MD 21250}

\altaffiltext{5}{Laboratory for High Energy Astrophysics, 
NASA's Goddard Space Flight Center, Code 662
Greenbelt, MD  20771.}

\altaffiltext{6}{School of Physics and Astronomy, Raymond and Beverly Sackler 
Faculty of Exact Sciences, Tel-Aviv University, Tel-Aviv 69978, Israel.}

\begin{abstract}

We present observations of the UV absorption lines in the Seyfert 1 galaxy 
NGC 3516, obtained at a resolution of $\lambda$/$\Delta\lambda$ $\approx$ 
40,000 with the Space Telescope Imaging Spectrograph (STIS) on 2000 October 
1. The UV continuum was $\sim$4 times lower than that observed during 1995 
with the Goddard High Resolution Spectrograph (GHRS), and the X-ray flux from 
a contemporaneous {\it Chandra X-ray Observatory (CXO)} observation was a 
factor of $\sim$8 below that observed with {\it ASCA}. The STIS spectra show 
kinematic components of absorption in Ly$\alpha$, C~IV, and N~V at radial 
velocities of $-$376, $-$183, and $-$36 km s$^{-1}$ (components 1, 2, and 
3$+$4, respectively), which were detected in the earlier GHRS spectra; the 
last of these is a blend of two GHRS components that have increased greatly 
in column density. Four additional absorption components have appeared in the 
STIS spectra at radial velocities of $-$692, $-$837, $-$994, and $-$1372 km 
s$^{-1}$ (components 5 through 8); these may also have been present in 
earlier low-flux states observed by the {\it International Ultraviolet 
Explorer (IUE)}. Based on photoionization models, we suggest that the 
components are 
arranged in increasing radial distance in the order, 3$+$4, 2, 1, followed
by components 5 -- 8. We have achieved an acceptable fit to the X-ray data 
using the combined X-ray opacity of the UV components 1, 2 and 3$+$4.
By increasing the UV and X-ray fluxes of these models to match the previous 
high states, we are able to match the GHRS C~IV column densities, absence of 
detectable C~IV absorption in components 5 through 8, and the 1994 {\it ASCA} 
spectrum. We conclude that variability of the UV and X-ray absorption in NGC 
3516 is primarily due to changes in the ionizing flux.

\end{abstract}


\section{Introduction}

It is now understood that intrinsic absorption is a common phenomenon in
Seyfert 1 galaxies,
present in more than half of those observed with the spectrographs 
aboard the {\it Hubble Space Telescope (HST)} (Crenshaw
et al. 1999). Among those Seyferts that show absorption, high ionization 
resonance lines
such as N~V $\lambda\lambda$1238.8, 1242.8
and C~IV $\lambda\lambda$1548.2, 1550.8 are always present, along with
Ly$\alpha$, while lower ionization lines, such as 
Si~IV $\lambda\lambda$1393.8, 1402.8, and Mg~II
$\lambda\lambda$2796.3, 2803.5, are less common.
The absorption lines can be 
blueshifted (by up to 2100 km s$^{-1}$) with respect to the
systemic velocities of the host galaxies, indicating net radial outflow. 
However, UV absorption lines can arise in dusty gas within the plane of the
host galaxy (Crenshaw et al. 2001; Crenshaw et al. 2002); these are often
saturated and at velocities close to systemic.
The ionic columns can be highly variable, which may be the result
of changes in response to the ionizing continuum, as first suggested 
in case of NGC 4151 by 
Bromage et al. (1985) and Espey et al. (1998), and discussed in
detail in Crenshaw et al. (2000a) and  Kraemer et al. (2001a), or 
transverse motion, as appears to be the case for NGC 5548 and NGC 3783 
(Crenshaw \& Kraemer 1999; Kraemer, Crenshaw, \& Gabel 2001b). In fact, 
both mechanisms may be at work in individual AGN. In either case, the 
variability is indicative of the promixity of the absorbers to the central 
active nucleus of these galaxies; in the case of NGC 4151, constraints on the
density and ionization of the strongest absorption system place it within 
0.03 pc of the continuum source (Kraemer et al. 2001a).

The presence of intrinsic absorption, typically in the form of 
bound-free edges of O~VII and O~VIII, has been detected in the X-ray 
spectra of a similar fraction of Seyfert 1 galaxies (Reynolds 1997; 
George et al. 1998). Most recently, spectra obtained with the {\it Chandra
X-ray Observatory (CXO)}
have revealed that X-ray absorption lines associated with this
material are also blue-shifted (Kaastra et al. 2000; Kaspi et al. 2000,
Collinge et al. 2001).
The connection between the X-ray and UV absorption is complex and
there is likely a range of physical conditions within the absorbers (see 
Kriss et al.
1996a; Kriss et al. 2000; Krolik \& Kriss 2001). Furthermore, there
is evidence that the X-ray absorbers in some sources must also
be multi-zoned (Otani et al. 1996; Reynolds et al. 1997).

NGC 3516 ($z$ $=$ 0.00875) is one of the few Seyfert 1 galaxies
with UV absorption lines, specifically N~V, C~IV, and Si~IV, strong enough to 
have been detected with the {\it International Ultraviolet Explorer (IUE)} (
Ulrich \& Boisson 1983). These lines are
known to vary on timescales as short as weeks (Voit, Shull, \& Begelman 1987:
Walter et al. 1990; Kolman, Halpern, \& Martin 1993). Walter et al. suggested
that the C~IV absorption consists of a narrow, apparently stable component
in the core of the broad emission lines and a variable, broad, blue-shifted
component. Koratkar et al. (1996) found the 
blueshifted component present in {\it IUE} spectra in 1978, but it had disappeared between
1989 and 1993, and has been absent in subsequent {\it Hopkins Ultraviolet
Telescope} (HUT) (Kriss et al. 1996b) and
{\it HST}/Goddard High Resolution Spectrograph (GHRS) (Crenshaw, Maran, \& Mushotzky 1998) observations. Based on 
GHRS spectra, Crenshaw et al. (1998)
determined that the absorption in the core of the line (Walter et al.'s
``stable'' component) consisted of two
broad components, with velocities of $-$380 
and $-$ 150 km s$^{-1}$, and two narrow components at $-$90 and
$-$30 km s$^{-1}$.

  NGC 3516 also exhibits strong, variable X-ray absorption (Kolman et al. 1993;
Nandra \& Pounds 1994; Kriss et al. 1996a; Mathur, Wilkes, \& Aldcroft 1997). 
Although Mathur et al. (1997) suggested that the X-ray and UV absorption 
arises in 
gas characterized by a single set of physical conditions, Kriss et al. (1996a) modeled 
the X-ray absorber as two zones, one with an ionization parameter (the ratio 
of the density of photons with energies $\geq$ 13.6 eV
to the number density of hydrogen atoms at the illuminated face of
the zone) U $=$
1.66 and a total hydrogen column density of 1.4 x 10$^{22}$ cm$^{-2}$ and a lower
ionization zone with U $=$ 0.32 and a total column density of 6.9 x 
10$^{21}$ cm$^{-2}$, and argued that not all of the UV lines could form in 
these zones. In our previous paper (Netzer et al. 2002; hereafter Paper I), we presented
an analysis of the {\it CXO}/Low Energy Transmission Grating Spectrograph
(LETGS) spectrum obtained on 2000 October 6. 
The X-ray continuum flux, Flux$_{6.4 KeV}$
$=$ (1.6 $\pm$ 0.2) x 10$^{-4}$ photons cm$^{-2}$ s$^{-1}$ KeV$^{-1}$, was 
close to its historic minimum, roughly a factor
of 8 -- 10 below maximum (observed in 1994 with {\it ASCA}). 
We found that the variations in observed flux and spectra at these epochs is
consistent with an absorber of constant column density (7.9 x 10$^{21}$ 
cm$^{-2}$)
whose ionization state has changed in response to the variation
in the ionizing continuum, specifically: for the current low flux state, 
U $=$ 0.19, while it would be a factor of 8 to 10 higher
in the high flux state (depending on the amplitude of the UV continuum 
variations), in rough agreement with the high ionization component described
by Kriss et al. (1996a). The model predicts 
large column densities ($\sim$ 10$^{17}$ cm$^{-2}$) for
N~V and C~IV, hence one or more of the observed kinematic components of
the UV absorption should be associated with the X-ray absorber.

We obtained echelle spectra of NGC 3516 with the Space Telescope Imaging Spectrograph
(STIS) on HST under a STIS Guaranteed Time Observer Program. In this paper,
we present our analysis of the physical conditions in the UV absorbers,
nature of their variability, and the connection to the X-ray absorption detected
in our contemporaneous {\it CXO}/LETGS observation. 

\section{Observations and Data Analysis}

\subsection{New STIS and Previous GHRS Observations}

We obtained STIS echelle spectra of the nucleus of NGC 3516
through a 0\arcsecpoint2 x 0\arcsecpoint2 aperture on 2000 October 1. These 
observations and two previous observations of the C~IV region with 
the Goddard High Resolution Spectrograph (GHRS) are detailed in Table 1. The 
reduction of the GHRS spectra is described in Crenshaw et al. (1998, 1999). 
We reduced the STIS echelle spectra using the IDL software developed at NASA's 
Goddard Space Flight Center for the STIS Instrument Definition Team.
The data reduction includes a procedure to remove the background light from 
each order using a scattered light model devised by Lindler (1999). The 
individual orders in each echelle spectrum were spliced together in the 
regions of overlap.

Figure 1 compares the C~IV regions from the high-resolution spectra obtained 
at the three epochs. As noted in Crenshaw et al. (1998, 1999), the GHRS 
observations in 1995 showed four kinematic components of absorption in C~IV
(which we will refer to as the ``core components''). 
Although there was a decrease in the underlying continuum plus broad 
C~IV emission over the six-month interval in 1995, there was no evidence for a 
change in the ionic column density (or radial velocity coverage) of any 
component. The new STIS observations in 2000 October show a further
decrease in the continuum plus broad C~IV emission, plus a dramatic change in 
some of the C~IV absorption components. In the STIS spectrum, there is a 
strong absorption feature at the position of the weak GHRS components 3 and 4. 
In addition, there is 
significant new absorption in the blue wing of the broad C~IV profle, in the 
1553 -- 1559 \AA\ region. This new absorption is reminiscent of the broad, 
variable blueshifted absorption seen in the C~IV profile from early {\it IUE} 
observations (Koratkar et al. 1996), which will be discussed later in this 
section. There is no evidence for significant changes in components 1 and 2, 
which are heavily saturated at all three epochs.

Figure 2 shows portions of the STIS E140M spectra where intrinsic absorption 
is detected in other ions. Fluxes are plotted as a function of the radial 
velocity (of the strongest member, for doublets), relative to the systemic 
redshift of z $=$ 0.00875 from the optical emission lines (Crenshaw et al. 
1999), since H~I 21-cm measurements are not available.
Each of the original 4 absorption components are present in Ly$\alpha$ and the 
doublets of N~V $\lambda\lambda$ 1238.8, 1242.2 and C~IV $\lambda$ 1548.2, 
1550.8. In addition, Si~IV $\lambda\lambda$ 1393.8, 1402.8 absorption is 
present in components 1 and 2. We have also detected C~II $\lambda$1334.5, 
C~II* $\lambda$1335.7 (fine-structure line), and Mg II $\lambda\lambda$ 
2796.3, 2803.5 intrinsic absorption, all in component 1 only. We 
have not reliably detected any other intrinsic absorption line in 
the E140M or E230M spectra.

We have separated the new blueshifted absorption into four distinct kinematic 
components (5 -- 8; which we will refer to as the ``blueshifted components''). Our criteria for identification of a component are that 
it must be distinct in radial velocity space (i.e., not severely blended with 
other components), and clearly present in N~V or C~IV. Component 
5 is by far the strongest and broadest new component, whereas components 6 -- 
8 are relatively weak.

\subsection{Measurements and Observational Results}

The procedures we used to measure the intrinsic absorption lines 
follow those of Crenshaw et al. (1999). To determine the shape of the 
underlying emission, we fit a cubic spline to regions on either side of the 
absorption, and then normalized the absorption profiles by dividing the 
observed spectra by the spline fits. 
We determined the covering factor C$_{los}$, which is the fraction 
of continuum plus emission that is occulted by the absorber in our line of 
sight, using the technique of Hamann et al. (1997) for doublets. 
We have relied on the C~IV doublets to determine these values, since many of 
the N~V doublets are contaminated by Galactic absorption or 
blending of kinematic components. Due to the modest 
signal-to-noise of the spectra and blending of components in the wings, 
C$_{los}$ can only be determined reliably in the cores of the lines.
We are not able to accurately deconvolve components 3 and 4 in the STIS 
spectra (see Figure 2), so they are treated as one component (3$+$4) in this 
paper.

Table 2 gives the radial velocity centroids, widths (FWHM), covering factors, 
and associated one-sigma errors for each kinematic component. 
These velocity centroids represent the means and standard deviations 
determined from the measurements of the C~IV, N~V, Si~IV, and Ly$\alpha$
profiles (when available for the latter two). Compared to the GHRS 
measurements (Crenshaw et al. 1999), there is no evidence for a change in the  
radial velocities of components 1 -- 4, except for perhaps a slight shift in 
component 2 from $-$148 km s$^{-1}$ (GHRS) to $-$183 km s$^{-1}$ (possibly due 
to increased absorption in the blue wing of this component). The radial 
velocity centroid of component 3$+$4 is very close to the position of the GHRS 
value for component 4, suggesting (but not proving) that component 4 may be 
the one responsible for the large increase in absorption. Our original 
suggestion that components 3 and 4 may be due to the interstellar medium or 
halo of the host galaxy (Crenshaw et al. 1998) is therefore still a possible 
explanation for one of these components (probably component 3), but not the 
other (our models are based on the assumption that these components
are essentially co-located, near the ionizing source).

To determine the FWHM of the components, we used values from relatively 
unsaturated lines, to get an indication of the true velocity spread for each 
component. Thus, we used the STIS Si~IV lines for components 1 and 2, the GHRS C~IV 
lines for components 3 and 4, and the STIS Ly$\alpha$, N~V, and C~IV lines for 
components 5 -- 8. For components 1 and 2, the FWHMs are therefore smaller 
than those measured from the GHRS C~IV profiles (Crenshaw et al. 1999), since 
the latter are heavily saturated.

Next, we examine the covering factors in Table 2. The covering factors for 
components 1 and 2 are {\it consistent} with a value of one, although we 
cannot rule out the possibility that they are as small as 0.96. 
Component 3$+$4 shows the strongest evidence for a non-unity covering factor. 
Although the aperture for the STIS observations is small, there is likely to 
be some emission from the narrow-line region (NLR) that is unocculted
(see Kraemer et al. 2001; Brotherton et al. 2002).
This could explain the nonuntity covering factor for component 3$+$4 since it 
is near the systemic redshift, where the NLR contribution is 
strongest. To illustrate this point, we have have simulated the NLR 
contribution by reproducing Gaussians at the systemic velocity in Figure 2.
We kept the doublet ratios fixed at 2:1, chose a single value for the FWHM
($=$ 165 km s$^{-1}$) that provided a good overall fit, and scaled the 
intensities to obtain the best fit for each ion. Given our assumptions, the 
NLR emission provides a reasonably accurate match to the troughs of the 
saturated lines (note the contamination of N~V $\lambda$1242.8 by Galactic 
S~II). The error in covering factor for component 5 is too large to make a definite 
conclusion. We could not determine reliable values for 
components 6 -- 8, since the profiles are shallow and relatively noisy. Lower 
limits from L$\alpha$ and estimates from N~V are consistent with the covering 
factors determined from C~IV. 

To determine the ionic column densities, we converted each normalized profile 
to optical depth as a function of radial velocity, and integrated across the 
profile, as described in Crenshaw et al. (1999). 
For components 1 and 2, the Si~IV profiles are much narrower than the profiles 
of Ly$\alpha$, N~V, and C~IV, suggesting that the latter are heavily saturated 
(assuming the same intrinsic velocity dispersion for all ions). By the same 
token, components 3 and 4 in the GHRS spectra are narrow in C~IV compared to 
the STIS component 3$+$4, again suggesting heavy saturation in the latter.
Furthermore, the possibility of nonunity covering factors for components 1, 2, 
and 3$+$4 suggests that the column densities may be much larger than would 
be determined for C$_{los}$ = 1. Thus, we can only determine lower limits to 
the column densties for these lines (by assuming C$_{los}$ = 1), and the 
column densities for components 1 and 2 from the GHRS spectra (Crenshaw et al. 
1998, 1999) should be taken as lower limits. For component 5, even if 
C$_{los}$ = 0.6, the depth of the lines are such that they are not heavily 
saturated. As we mentioned previously, we do not have reliable measurements of 
C$_{los}$ for components 6 -- 8. Since these components are shallow, 
they are unlikely to be saturated (including Ly$\alpha$ in these components), 
unless the covering factors are extremely small, which seems unlikely. Thus, 
we have assumed C$_{los}$ $=$ 1 for components 5 -- 8.

Table 3 gives our measurements of STIS and GHRS column densities for all lines 
that have been detected in at least one component. We include lower 
limits for heavily saturated lines and upper limits for undetected lines
to constrain the photoionization models.

\subsection{Comparison with Previous Low-Resolution Spectra}

To pursue a possible connection between the continuum flux and the 
strength of the C~IV absorption, we have examined all of the far-UV spectra 
that have been obtained of NGC 3516. Table 4 lists the additional far-UV 
spectra (i.e., those not listed in Table 1), which were obtained at lower 
spectral resolutions by the 
{\it IUE}, {\it HUT}, and the Faint Object Spectrograph (FOS) and Space 
Telescope Imaging Spectrograph (STIS) on {\it HST}. We have retrieved the most 
recently processed versions of these spectra from the {\it IUE} and {\it HST} 
archives in order to measure their continuum fluxes and examine their C~IV 
profiles. For the {\it HUT} spectrum, we determined the continuum flux 
directly from the plot in Kriss et al. (1996b).

We measured the continuum fluxes by averaging the points in a bin centered at 
1460~\AA\ (observed frame) with a width of 20 \AA. We determined the one-sigma 
flux errors from the standard deviations of the averages; for the {\it IUE} 
spectra, this technique is known to overestimate the errors (Clavel et al. 
1991), and we 
scaled the errors for these observations by a factor of 0.7 to ensure that 
observations taken on the same day agreed to within the errors (on average).
For the two GHRS spectra, the continuum regions adjacent to the broad C~IV 
emission profile were not observed. In this case, we used separate fits to the 
continuum and broad emission in the STIS C~IV region and tried different 
linear combinations of these fits until an accurate match was obtained to the 
continuum plus broad emission in each GHRS spectrum.

Figure 3 shows that the far-UV continuum of NGC~3516 has varied 
dramatically over a span of 22 years. Early {\it IUE} observations were 
obtained when NGC 3516 was in low to moderate states, whereas later 
observations, including the intensive monitoring campaign of 1996 ($\sim$JD 
2,490,000, see Koratkar et al. 1996), were obtained primarily at a high state. 
The last {\it IUE} observations and subsequent observations by other 
instruments demonstrate the rapid variability of NGC 3516; in particular, the 
FOS observations show a factor of 5 variation (ratio of brightest to faintest) 
over the course of 11 months (see Goad et al. 1999). However, the trend is to 
lower flux levels in the most recent obervations. The new STIS observations 
were obtained at very low (but not unprecedented) continuum flux levels.

To examine the previous behavior of the C~IV profile as a function of 
continuum luminosity, we have placed the {\it IUE} spectra into three 
categories based on their continuum fluxes at 1460 \AA: low (0 -- 2.0 x 
10$^{-14}$ ergs s$^{-1}$ cm$^{-2}$ \AA$^{-1}$), medium (2.0 -- 4.0 x 
10$^{-14}$ ergs s$^{-1}$ cm$^{-2}$ \AA$^{-1}$), and high (4.0 -- 7.0 x 
10$^{-14}$ ergs s$^{-1}$ cm$^{-2}$ \AA$^{-1}$). We averaged the spectra 
(weighting by exposure time) in each category. Figure 4 shows the average 
spectra in the C~IV profiles. For comparison, we binned the GHRS and STIS 
spectra to the same wavelength sampling as the {\it IUE} spectra (1.67 \AA\ 
per bin), and smoothed slightly to match the {\it IUE} resolution (6~\AA\ 
FWHM). We overplot the binned high-state GHRS and low-state STIS spectra in 
Figure 4.

The {\it IUE} spectra in Figure 4 show a remarkable change in the C~IV profile 
as a function of continuum luminosity. The blue emission peak just shortward 
(in wavelength) of the central absorption feature is stronger than the red 
peak at high luminosity, but becomes weaker and shifts to shorter wavelength 
at lower luminosities. The binned GHRS and STIS spectra show the same effect. 
Comparison with Figure 1 shows that this effect is due to increased absorption 
from components 5 -- 8 in the low-state STIS spectrum. This effect has been 
noticed in previous {\it IUE} studies. Walter et al. (1990) characterized the C~IV 
absorption in {\it IUE} spectra as a blend of a narrow stable component in the 
core of the emission line plus a variable broad and blueshifted component. The 
similar appearances of the low-state {\it IUE} and STIS C~IV profiles indicate 
that the variable blue-shifted absorption covers a similar range in radial 
velocity, and {\it suggest} that they originate from the same kinematic 
components. This would not be too surprising, since long-term stability in the 
radial velocities of absorption components is a common trait in Seyfert 1 
galaxies (Crenshaw et al. 1999).

Koratkar et al. (1996) showed that the blueshifted component disappeared in 
{\it IUE} spectra between 1989 October and 1993 February (when NGC 3516 was in 
a high state). At the time, it was not clear if this effect was due to 
evolution of the absorbers (e.g., bulk motion of gas out of the line of sight) 
or to a change in the ionization of the gas (e.g., a decrease in the C~IV 
column due to an increase in the ionizing continuum, coincident with the 
increase in the UV continuum). There are several pieces of evidence that 
indicate that the latter (variable ionization) is certainly an important, and 
probably dominant, contributor to the C~IV absorption variability. 1) Our 
examination of individual {\it IUE} spectra indicate that the C~IV profile 
variability is more correlated with continuum flux than with time. In 
particular, the moderately high states at early times and the moderately low 
states at the end of the {\it IUE} observations confirm the above trend. 2) 
Walter et al. (1990) found an anticorrelation between continuum flux and C~IV 
equivalent width in the {\it IUE} spectra, consistent with this scenario. 3) 
The individual FOS observations, which span a large range (factor of 5) in 
continuum flux, show the same effect of decreasing blue peak 
with decreasing flux (see the spectra in Goad et al. 1999). 4) The high 
resolution STIS and GHRS spectra confirm this effect, and identify the 
kinematic components responsible for the change in the C~IV profile. We note 
that the same effect has been seen in the intrinsic O~VI absorption in 
NGC~3516 (Hutchings et al. 2001): high-velocity absorption (consistent with 
components 5 -- 8) appeared in a low-state {\it FUSE} observation, but was not 
apparent in a moderately high-state {\it HUT} observation obtained five years 
earlier. Thus, we believe we have strong evidence for variable ionization of 
the absorption in NGC~3516, such that the high-velocity UV absorption 
disappears at high continuum fluxes.

\section{Photoionization Models}

\subsection{Inputs to the Models}

 Photoionization models for this study were generated using the
 code CLOUDY90 (Ferland et al. 1998). We have modeled the absorbers
 as matter-bounded slabs of atomic gas, irradiated
 by the ionizing continuum radiation emitted by the central source.
 As per convention, the models
 are parameterized in terms of the ionization parameter,
 U. Each separate kinematic component was initially modeled with one
 set of initial conditions, i.e., U, n$_{H}$, and the total
 hydrogen column density, N$_{H}$ ($=$ N$_{H~I}$ $+$ N$_{H~II})$\footnote{We 
 use N$_{XM}$ to denote the ionic column density, where ``X'' is the atomic 
 symbol and ``M'' is the ionization state.}. 
 For the models, we have assumed only thermal 
 broadening, since 1) the absorption line widths could be due to the
 superposition of unresolved kinematic components and 2) comparison models 
 assuming turbulent velocities
 of $\leq$ 300 km s$^{-1}$ predict nearly identical ionic columns.
 Since the strongest absorption lines in components 1, 2, and 3$+$4
 (Ly$\alpha$, N~V, and C~IV) are
 saturated, we can only compare the model predictions to lower limits
 for the ionic 
 column densities. However, the predictions for these 
 components are constrained by the upper limits, or relatively small column 
 densities, derived for lower ionization species, such as C~II and Mg~II.
 Furthermore, total X-ray opacity predicted by the UV absorbers models cannot
 exceed that which we determined from the {\it CXO}/LETGS spectrum (Paper I).

 In Paper I, we suggested that the total reddening in NGC 3516 may be slightly
 higher than E$_{B-V}$ $=$ 0.06 (determined by Kriss et al. [1996b]).
 First, using combined STIS and FOS data from similar flux states,
 we dereddening the continuum using the Galactic curve (Savage \& Mathis 1979) until
 we matched the NGC 4151 continuum. With the caveat that the NGC 3516 spectrum 
 is from a 
 low state, hence the FUV could be
 intrinsically depressed (Clavel et al. 1991; Crenshaw et al. 1996), while the NGC 4151 spectrum is from a 
 moderately high state, we derived an upper limit 
 of E$_{B-V}$ $\leq$ 0.15. Alternately, we found, using the average continuum
 fluxes at 1460\AA~ and 5510\AA~ determined by Edelson et al. (2000) during 
 their STIS monitoring campaign that an
 underlying continuum of the form F$_{\nu}$ $\propto$ $\nu^{\alpha}$, where
 $\alpha = -1$, is 
 consistent
 with an E$_{B-V}$ $\approx$ 0.10. Therefore, we have adopted a reddening
 value of E$_{B-V}$ $=$ 0.10 $\pm$ 0.05, of which 0.04 is
 Galactic (Schlegel, Finkbeiner, \& Davis 1998). This is in rough agreement with
 Kriss et al. (1996b), but suggests an UV power-law that is more consistent 
 with typical values for low luminosity AGN (Wills et al. 1985). If the dust 
 composition and dust-to-gas ratio are the same as in our Galaxy (see Mathis,
 Rumpl, \& Nordsieck 1977), the intrinsic reddening, E$_{B-V}$ $\approx$ 0.06,
 requires a neutral hydrogen column of $\sim$ 3.0 x 10$^{20}$ cm$^{-2}$ (Shull
 \& van Steenberg 1985). Note,
 however, that the gas within which the dust resides need not be neutral
 (see Brandt, Fabian, \& Pounds 1996), nor might we expect that the
 dust near an AGN is the same as that found in the Galactic ISM 
 (Pitman, Clayton, \& Gordon 2000; Kraemer et al. 2000; Crenshaw et al. 
 2002).

 As in Paper I, we have assumed that the UV power-law continuum extends out 
 into the EUV, at which point it must steepen to meet the low-energy extension
 of the X-ray continuum.  Hence, we have approximated the spectral energy
 distribution (SED), as a power-law, with the following indices:
 $\alpha = -1$ below 100~eV, $\alpha = -2.3$ over the range 100 ~eV 
 $\leq h\nu <$ 500~eV, and $\alpha = -0.7$ above 500~eV. 
 Assuming H$_{0}$ $=$ 75 ks s$^{-1}$ Mpc$^{-1}$, the total 
 Lyman continuum photon luminosity is Q $\approx$ 2 x 10$^{53}$ photons 
 sec$^{-1}$, roughly twice that for NGC 4151 when it was in a recent low 
 luminosity state (Kraemer et al. 2001a).
 It should be noted that the
 power-law break could occur at lower energies, in which case the ratio of the
 X-ray ionization parameter (see George et al. 1998) to U would be somewhat
 larger, and the relative fractions of highly ionized species, such as O~VII 
 and O~VIII, would be somewhat higher than our predictions. 
 We have also modeled the UV absorbers in the 1995 high state observed 
 with the GHRS. In addition to a higher overall
 ionizing luminosity, we have assumed a slightly different SED, as discussed
 in Section 4.3.

 We have assumed roughly solar element abundances (cf. Grevesse \& Anders 1989),
 which are, by number relative to H, as follows: He $=$ 0.1, 
 C $=$ 3.4 x 10$^{-4}$, N $=$ 1.2 x 10$^{-4}$, O $=$ 6.8 x 10$^{-4}$,
 Ne $=$ 1.1 x 10$^{-4}$, Mg $=$ 3.3 x 10$^{-5}$, Si $=$ 3.1 x 10$^{-5}$,
 S $=$ 1.5 x 10$^{-5}$, and Fe $=$ 4.0 x 10$^{-5}$. The absorbing gas
 is assumed to be free of cosmic dust.

\subsection{Relative Distances of the Absorbers}

 As discussed in Section 2, the covering factor for each of the core
 components of absorption is $>$ 0.9. Hence, the simplest
 geometric arrangement has the absorbers separated in radial position from the
 central source, but subtending the same solid angle, i.e., ``screening''
 each other from the ionizing continuum. As a result, the ionizing continuum
 that reaches the absorbers is diluted by the inverse square of
 its distance and filtered by intervening gas. This arrangement has
 important consequences for the response of the screened absorbers to changes
 in the luminosity of the central source, since they are affected by both
 changes in the ionizing flux {\it and} the opacity of the
 intervening gas.   

 The core components were present in C~IV 
 in GHRS spectra taken while the continuum flux of NGC 3516 was $\geq$ 4
 times higher than during our STIS observations, and 
 components 1 and 2 were highly saturated (Crenshaw et al. 1998). 
 In the STIS spectra each of the core components possesses saturated Ly$\alpha$, N~V, and 
 C~IV lines. Component 1
 has saturated Si~IV and lines from several low ionization species, hence
 must be the most optically thick at the He~II Lyman limit. Based on 
 our photoionization models, the most dramatic changes in the 
 transmitted ionizing continuum occur when there is a large drop
 in the He~II opacity, to the extent that a screened component 
 which possessed a C~IV column of $>$ 10$^{15}$ cm$^{-2}$ 
 (e.g., a model with N$_{H}$ $=$ 10$^{21}$ cm$^{-2}$ and a 
 low-state ionization 
 parameter U $=$ 0.1), would have essentially all of its carbon 
 ionized out of C~IV during the high state.
 Since each of these absorbers was 
 present while NGC 3516 was in a high flux state, it is unlikely that 
 component 1 is closest to the ionizing source. Expanding on this, we
 found that the best fit to the observed ionic columns 
 is achieved when the three components are arranged in increasing
 order of He~II optical thickness, specifically: component 3$+$4 lying
 closest, followed by 2, then 1. Similarly, the absence of the blueshifted
 components during high flux states of NGC 3516 is consistent with
 them being
 screened by gas that is optically thick during low flux states, and 
 hence we chose
 to have them positioned outside the core components.
 
 We have attempted to model each kinematic component as a single zone, 
 adding additional zones only when necessary. As typical in such analyses, 
 the resulting model parameters are a possible, but not necessarily unique,
 solution set. However, the dual constraints of screening and the total
 X-ray opacity significantly reduce the parameter space
 for the models. Finally,
 the true test for this geometry and physical parameters is whether such a 
 model can 
 replicate the high-state column densities, as we discuss in Section 4.3.

\section{Model Results}

\subsection{The UV Absorbers}

 The model parameters for each component, including U, N$_{H}$, and
$<T_{e}>$ (the mean electron temperature),
are listed in Table 5. Implicit in our assumptions is that the distances 
between these components 
are significantly smaller than their average radial distance, i.e, 
they are roughly co-located.
Also, note that the 
predicted temperatures are sensitive to the choice of density and, hence, 
should be considered as very rough approximations.

The predicted ionic columns for each component are compared to the measured 
values
in Table 6 (for the full set of predicted column densities for the core
components, see Tables 7 
through 10).
For component 3$+$4, N$_{H}$ and U were constrained by the lower limits
to N$_{H~I}$ N$_{C~IV}$ and N$_{N~V}$, and the upper limit to N$_{Si~IV}$
(see Table 3). Component 2 is ionized by the continuum transmitted by
3$+$4, and is constrained by the value of N$_{Si~IV}$, as well as the
aforementioned lower limits. Both of these components were successfully modeled
as single-zoned absorbers. 

For component 1, a single-zoned model underpredicts C~II, while 
overpredicting Mg~II. This is a result of the ionization structure
of the slab interior to the point where the gas becomes optically
thick at the He~II Lyman limit (54.4 eV). At this point most of 
the magnesium is in the form of Mg~III, which has an ionization potential of 
80 eV, little of which can be ionized to Mg~IV by the heavily absorbed 
ionizing continuum. On the other hand, most of the carbon is C~IV, since 
C~III has an ionization potential of 47.7 eV, i.e., below that of 
He~II. Hence, while Mg~II is populated by recombination of Mg~III, there is 
little C~II, due to the relatively small fraction of carbon in the form of 
C~III. If the ionization parameter is significantly lower, the
higher ionization states are less populated due to the paucity of 
ionizing photons, and the situation is mitigated, in the sense that
the C~III and Mg~III zones overlap more completely. Hence, we used a 
two-zoned model to fit component 1. The
choice of parameters for the two components is somewhat arbitrary, however,
for the sake of simplicity, we assumed that the N~V, C~IV, and 
Si~IV arise in one component (1A), while the C~II and Mg~II are formed in
the other (1B). Note
that, while the prediction for N$_{C~II}$ is quite good, the model 
overpredicts
N$_{Mg~II}$ by a factor of $\sim$ 3.5. This may be the consequence of 
our assumption of solar abundances, as we discuss in Section 5.2.  

Interestingly, components 1A and 1B predict a combined neutral column of
N$_{H~I}$ $\approx$ 10$^{17}$ cm$^{-2}$, which is less than half the column
determined from the 1995 HUT spectrum (Kriss et al. 1996a). Part of the
difference may be due to the contribution from component 2, since these
components are not separable at low resolution. However, the continuum
flux observed by HUT was roughly a factor of two higher
than in the STIS spectra and, as we discuss in Section 4.2, the
total neutral column should, therefore, have been lower. Nevertheless, we do not
detect any additional low ionization lines in the STIS spectra that might be 
associated 
with an unmodeled neutral component. This apparent discrepancy may be
due to the presence of a small additional column of neutral gas within 
our line-of-sight during the time of the HUT observations.

As shown in Table 3, the N$_{N~V}$/N$_{H~I}$ and N$_{C~IV}$/N$_{H~I}$ ratios 
for the blueshifted components are unusual 
in the sense that both are large. Although H~I columns can be
quite small in very highly ionized gas (U $>$ 0.5), the fractional populations
of the lithium-like ionic states tend to be similary small. In part, the
unusual ionization balance in these components may be the result of illumination
by a heavily absorbed continuum, which has very few photons with
energies between 54.4 eV and 100 eV (see Figure 5). In this case,
N~V is produced by X-ray ionization of N~IV by photons with energies
much greater than its ionization potential; the photoionization cross-section
at these energies is much lower than at the threshold energy. To overcome this
inefficiency, the ionization parameter must be high ($>$ 0.5), which in turn
ionizes a greater fraction of the hydrogen than in gas exposed
to an unfiltered continuum. 

While the N$_{N~V}$/N$_{C~IV}$ ratios for the blueshifted components are 
well-fitted by the models, N$_{H~I}$ is 
underpredicted for component 5, and overpredicted for the other components.
It must be noted that the overpredictions are {\it much} worse when an 
unfiltered continuum is used (we did not include the effects of continuum 
filtering between these components, but they are all optically thin at the
He~II Lyman limit, so the effect
is negligible). Somewhat more curious, given the casual similarity of these 
absorption features, is the large range in N$_{H}$ and U (see Table 5). 
However, these components experienced an extreme drop in the ionizing
flux, due to the drop in the intrinsic
continuum luminosity and the increase in opacity of the intervening gas,
over which time they may have undergone large temperature changes (see Section
5.2). It is possible that the apparently peculiar ionization balance in these 
components is a result of this recent history. Therefore, we only list the
predicted columns for component 5, which is the lowest ionization and
most thermally stable (which may explain why the predicted column 
densities are the most accurate).

\subsection{Origin of the X-ray Absorption}

Our models for the core components predict significant columns densities for
C~VI, N~VI, O~VI, and O~VII, hence, significant X-ray opacities.
As shown in Figure 5, the large He~II column supresses the soft-Xray flux.
Finally, the total column density for these components is N$_{H}$ $=$
7.7 x 10$^{21}$ cm$^{-2}$, which is nearly identical to the single zone model
described in Paper I. So, it is likely that the UV absorbers are responsible
for the X-ray absorption.

A full and detailed analysis of the X-ray 
spectra from NGC~3516 has already been presented in Paper~I.
We do not consider it necessary to repeat such an analysis here.
However, in order to test and illustrate the hypothesis that the 
UV absorbers and X-ray absorbers are dominated by the same 
components, we have performed direct comparisons of the 
models and data.
The ionic column densities for each of the 
core components (see Tables 7 -- 10) 
were used to predict the bound-free absorption edges in the 
X-ray band using the analytic expressions of
Verner \& Yakovlev (1995) and Verner et al. (1997). We have excluded
the blueshifted components since we have not constrained their physical states,
although it is our belief that their column densities are too low 
to contribute much to the X-ray absorption.
These were compared to the X-ray data assuming a Galactic
column density of $2.9\times10^{20}\ {\rm cm^{-2}}$
appropriate for this line of sight (Murphy et al 1996).

It should be stressed that this direct and simple comparison 
does {\it not} include the additional spectral components 
discussed in the more detailed treatment of Paper~I. 
Specifically, here we do not include the 
Compton ``reflection'' from optically-thick
material which is likely to dominate the observed X-ray
spectrum $\gtrsim 5$~keV, or 
or any emission from any other
circumnuclear material.

Nevertheless assuming the SED described in Section 3.1,
we find our column densities for the low-state
spectrum produce a predicted X-ray spectrum
similar to that of Paper I. Thus our model provides a comparable (acceptable)
description of the {\it CXO}/LEGTS data to that shown in Paper I, and
is strong evidence that the UV and X-ray absorption arise in the same gas, as 
suggested by Mathur et al. (1997), although the absorber consists
of multiple zones, characterized by different ionization parameters and 
column densities, as discussed in Kriss et al. (1996a). 
Also interesting is the lack of evidence for highly ionized 
gas (e.g., U $\sim$ 1, N$_{H}$ $>$ 10$^{21}$ cm$^{-2}$) within the 
line-of-sight to the active nucleus in NGC 3516, in contrast to 
NGC 3783 (Kaspi et al. 2000), NGC 5548 (Kaastra et al. 2000), and
NGC 4051 (Collinge et al. 2001). 

\subsection{Variability of the Absorbers}

In Paper I, we argued that the differences in the X-ray properties of NGC 3516
in the {\it CXO}/LETGS spectrum compared to earlier {\it ASCA} spectra were
due to variations in the continuum flux and the corresponding changes in
the ionization state and, therefore, opacity of the intervening absorber.
Since we have established that the UV and X-ray absorption occur in the 
same gas, it follows that the changes in UV absorbers must also 
be in response to variations in ionizing flux, rather than variations in 
total column density. As noted above, 
the continuum flux at 1460 \AA~ was roughly 4.5 times
greater in the GHRS observations than during the STIS GTO observations,
and represents the AGN in a high state. We assumed that the flux at the 
Lyman limit scaled by the same factor.
As there were no contemporaneous X-ray observations, we assumed that the 0.5 
keV flux increased by a factor of 9, as suggested in Paper I. Finally,
we assumed that the $\alpha$ remained $=$ $-$1.0 between 13.6 eV and 100 eV, 
which requires a change in $\alpha$ to $-$1.87 from 100 eV to 500 eV (although,
clearly, we do not know how the SED may have changed in the EUV to soft X-ray
band). Through an iterative analysis of the 1994 {\it ASCA} spectrum, we 
set the 0.5 
to 10 keV spectral index to $-$0.89.

In modeling the absorbers we held the column densities fixed and assumed the
same relative positions. The high-state continuum increased Q , and thus
U for the unscreened gas, by a factor of $\approx$ 4.6.
The predicted values of N$_{C~IV}$ for the core components are given in
Table 6, and match the GHRS values well, although the prediction for
component 2, while still indicative of saturation, is slightly below the 
lower limit (which may be indicative of a somewhat larger column density
in the earlier epoch, as suggested by the limits on N$_{H~I}$ determined
from the HUT spectrum [Kriss et al. 1996a]). Interestingly, in the low-state models, all the C~IV associated
with component 1 was from the higher ionization zone, while in the high-state
both zones of component 1 contribute to it. 
The high-state model for component 5 predicts a 
negligibly small value for N$_{C~IV}$, so the absence of the blueshifted
components can also be explained by the higher ionizing flux.

As with the low-state models,
we have taken the predicted columns densities (see Tables 11 -- 14) and 
used them to model the {\it ASCA} data obtained in 1994 April 
(e.g. George et al. 1998) when the nucleus of NGC~3516 was a factor
of 8 to 10 (at $\sim 6$~keV) brighter
(and the Compton "reflected" continuum less important).
We followed the standard procedures for the reduction and
analysis of {\it ASCA} data, ignoring the 5--7~keV band
(where there is emission from Fe K$\alpha$),
and fitting the time-averaged spectral
data from all four {\it ASCA} detectors simultaneously.
Only the spectral index and normalization of the
underlying power law continuum was allowed to vary
during the analysis
(the bound-free absorption appropriate
for the "high-state" column densities
were fixed at the values predicted by the UV models).
As expected, the index of the continuum
was in good agreement with our assumptions concerning the SED
($1.87 \lesssim \Gamma \lesssim 1.89$ at 90\% confidence).
The resulting fit is shown
in Figure 6, and it again matches the overall spectral characteristics.
Given the source is variable in both the
X-ray band, and in the overall shape of the SED,
and that there are no
simultaneous high resolution UV measurements, we consider our predictions to 
be in remarkable agreement with the X-ray observations
at this epoch.

\section{Discussion}

\subsection{Geometry of the Circumnuclear Gas}

We have shown that our model for the UV absorbers, with the components 
arranged in increasing radial distance and each covering the continuum source,
accurately predicts most of the ionic columns and the 
X-ray opacity. Given the success of the model, we may use it as
a starting point to unravel the geometry of the circumnuclear gas.
In paper I, based on the recombination timescale for 
O~VII, we placed a conservative lower limit on the density of X-ray absorber
of n$_{H}$ $\geq$ 2 x 10$^{6}$ cm$^{-3}$. 
For component 1B, we predict that N$_{C~III}$ $=$ 1.87 x 10$^{15}$ cm$^{-2}$.
The lack of detection of C~III$^{*}$ 1175 is indicative of N$_{C~III^{*}}$
$\lesssim$ a few x 10$^{13}$ cm$^{-2}$. For T $\sim$ 2.0 x 10$^{4}$K, this 
ratio of N$_{C~III^{*}}$/N$_{C~III}$ corresponds to a density of $\lesssim$
10$^{10}$ cm$^{-2}$ (Bromage et al.
1985; Kriss et al. 1992). If the high-ionization core components are
co-located with component 1b, their higher ionization parameters
require that they have densities of $\lesssim$ 10$^{8}$
cm$^{-3}$. Alternatively, for component 1A, we predict that N$_{C~III}$ 
$=$ 9.43 x 10$^{15}$ cm$^{-2}$. Based on the upper limit for N$_{C~III^{*}}$, the
density of component 1A is $<$ 10$^{9.5}$ cm$^{-3}$. Hence, we derive a loose 
upper limit of
n$_{H}$ $\lesssim$ 10$^{9}$ cm$^{-3}$ for the density of the
high-ionization core components. Given our estimate of
the continuum luminosity and the constraints on the
densities of the absorbers, the core components must lie between 
5.4 x 10$^{16}$ cm and 1.2 x 10$^{18}$ cm from the central source. 
This range in radial distance
is consistent with full-covering of the broad C~IV emission region, which has 
a radius of $\sim$ 4.5 light-days (Koratkar et al. 1996). 
By comparison, Kraemer et al. (2001a) estimated that the high density
(n$_{H}$ $\geq$ 10$^{9.5}$ cm$^{-3}$) UV absorber detected 
in NGC 4151, component D$+$E, must lie at a distance of $\lesssim$ 6.2 x 
10$^{16}$ cm from the central source (assuming H$_{0}$ $=$ 75 km s$^{-1}$ 
Mpc$^{-1}$). Hence, the radial distance of the absorbing gas in NGC 3516 is 
roughly consistent with that in NGC 4151, although possibly somewhat 
further from the central source and, apparently, less dense.
On average, the absorbers are in a much higher ionization state 
than component D$+$E, for which U $=$ 0.015. Since the ionizing
luminosity of NGC 3516 is only a factor of $\sim$ 2 greater than NGC 4151, 
observed in similar low-flux states, the greater ionization state
must be mostly due to the lower density. Another interesting difference is 
the weakness of the scattered light in the
troughs of the saturated absorption lines in NGC 3516: $<$ 5\% of the low-state
continuum flux, compared to $\geq$ 50\% of the low-state flux in NGC 4151 (Kraemer et al. 2001a).
For NGC 4151, we argued that the scattered light consisted of continuum $+$
broad-line emission,
reflected into our line-of-sight by free electrons in a component of highly 
ionized gas that extended beyond the
region covered by the absorbers. Either this component is weaker
in NGC 3516, the absorbers are covering more of the scattering region, or
the scattering region is further away from the nucleus (outside our
projected aperture).  

The observed emission-line fluxes can be used to constrain the global covering 
factor of the absorbers. The observed C~IV $\lambda$ 1550 flux is
$\approx$ 2.37 x 10$^{-12}$ ergs cm$^{-2}$ s$^{-1}$, or L$_{C~IV}$
$\approx$ 3.5 x 10$^{41}$ ergs s$^{-1}$. For a density of n$_{H}$ $=$ 10$^{8}$
cm$^{-3}$, the radial distance of component 3$+$4 is R $=$ 1.7 x 10 $^{17}$ cm
and the predicted C~IV flux at the illuminated surface of  
the 4 core component models absorbers is 1.5 x 10$^{6}$ 
ergs cm$^{-2}$ s$^{-1}$, hence, the covering factor must be $\leq$ 0.64. 
Interestingly, in its 
current low-flux state, approximately
50\% of the C~IV emission is from a component with FWHM of $\sim$ 1500
km s$^{-1}$ (see Figure 2 and Hutchings et al. [2001]), which is characteristic of the so-called intermediate line region 
(ILR; Wills et al. [1993]). If the absorbers are the source of the ILR
emission, the covering factor is $\sim$ 0.3. Note, that
although the core components cover both the broad and intermediate line contributions
to the C~IV emission, this does not imply that they must lie outside the
ILR. For example, self-absorption by the ILR gas could occur in an expanding 
shell (see Hutchings et al. 2001). If the absorption arises in the ILR, the 
ILR gas has a significant
radial velocity component, similar to the emission-line gas in the inner NLR 
of other Seyfert galaxies (Crenshaw
\& Kraemer 2000; Kaiser et al. 2000; Crenshaw et al. 2000b; Ruiz et al.
2001). A connection between the ILR and NLR components has been suggested
by Sulentic \& Marziani (1999), although our results suggest that the
ILR could extend to the outer edge of the BLR. 

\subsection{The Source of the Reddening}

From our estimate of the intrinsic reddening (see Section 4.1),
one would infer that there is a component of dusty gas of column density 
N$_{H}$ $\sim$
3 x 10$^{20}$ cm$^{-2}$ along our line-of-sight to the nucleus of NGC 3516.
Note that this is based on assumption of a Galactic dust/gas ratio; if the
ratio were lower, the column would be larger. As shown in Kraemer et al.
(2000) and Crenshaw et al. (2001), UV absorption lines can form in dusty 
gas surrounding an AGN. Typically, one would expect strong C~IV lines from
a column of this size. Hence, either the absorption lines associated with the
dusty column are buried in the troughs of the core components, or
the dust exists in one or more of these absorbers.

Following Barvainis (1987), we can estimate the dust sublimation radius,
i.e, the point interior to which dust grains would evaporate.
During its low-flux state, the ionizing luminosity of the central source
is $\sim$ 1.4 x 10$^{43}$ ergs s$^{-1}$. Assuming an average luminosity
a factor of 3 greater (see Section 2.3), and a dust sublimation temperature
of 1500 K (Salpeter 1977), the sublimation radius for NGC 3516 is 
$\lesssim$ 2.6 x 10$^{17}$ cm, or, roughly the same radial distance as the core
components. Although the absorbers are radially outflowing, we do not
know their starting point, or whether they were well-shielded from the
continuum source by intervening gas. Hence it is possible that some
grains have survived within them. Alternatively, some dust
could have been swept-up by the absorbers as they move outwards.
If, for example, the dust were within component 1, it may help
explain the overprediction of Mg~II (section 4.1). Furthermore, since
the column density of component 1 is roughly 5 times larger than our 
dusty column, the dust/gas ratio could be 20\% the Galactic ratio 
(indicative of the harsh conditions close to the active nucleus), and still
produce the observed reddening.
In any case, although it is possible the dust is in a heretofore
undetected component of gas, we suggest that the reddening may be due
to a small amount of dust within one or more of the core components.

\subsection{The Nature of the Outflow}

Perhaps the most unusual aspect of the intrinsic absorption in Seyfert galaxies
is the lack of evidence for radial acceleration.
In the case of NGC 4151, Weymann et al. (1997) attributed this to
the large radial distances of the absorbers. However, in Kraemer et al. (2001a)
we determined that component D$+$E was close enough to the central source
that it should undergo a detectable acceleration. To explain this, we 
suggested that
the large column density and optical thickness of D$+$E would make
radiative acceleration inefficient (see Williams 1972; Mathews 1974).
The physical conditions in the core components of NGC 3516 are similar to 
D$+$E,
hence the absence of detectable changes in radial velocities between
1995 and the present is understandable, although it is interesting that the blueshifted
components, which only appear during lowstates, appear at generally
the same velocities, and these are clearly optically thin
(of course, the radial acceleration
of the absorbers may simply be too small to detect in the $\sim$ 6 years between
the GHRS and STIS observations). Another
explanation is that the absorbers are not simply radially (or even
radiatively)
accelerated. Disk-wind models (e.g., Konigl \& Kartje 1994) require
that the gas in the outflow must first be levitated off the accretion disk. 
Bottorff, Korista, \& Shlosman (2000) have modeled the absorbers
as part of a magnetohydromagnetic wind, which has a strong
non-radial component. Elvis (2000) has also proposed an outflow scenario
that includes a non-radial component. 

If the absorption arises in 
disk-driven wind, the gas would have a strong transverse component of velocity. 
However our analysis (see also Paper I) suggests strongly that the 
variability of the absorption in NGC 3516 results from changes in ionization 
state due to the variable 
luminosity of the central source, rather than changes in column density due 
to transverse motion. One possibility is that the
individual kinematic components are associated with different flow
tubes or shells along our line-of-sight, which would originate at 
different radial positions at the base of the wind. Since the total
column density within the absorber does not appear to change, the amount of 
gas within these tubes must be fairly constant, although this does not imply
confinement but, rather, a continuous flow. Although the 
source 
luminosity changes dramatically, the flow density appears to be
independent of the short-term
changes in the flux, which may imply that it is not radiatively driven off the 
disk (see, again, Bottorff et al. 2000).

Interestingly, in our model the radial velocities of components 1 through 5 
increase with distance from the central source. This would be
expected if they each originated at the roughly the same distance, and have 
been continuously accelerated, e.g., via resonance line driving or some
other radiative mechanism. However, the velocities can also depend on the 
radial distance at which the different components originated, hence a range
in velocities may not be an indication of acceleration, but of different
``launch-pads'' for the components of the flow.

Another interesting aspect of the UV absorbers is their apparent stability; the
core components vary in ionization state but have shown evidence for
significant C~IV columns for two decades.
As discussed in Krolik, Mckee, \& Tarter (1981) and, more recently,
in Krolik \& Kriss (2001), there are two regimes
of ionization (characterized by the ``pressure'' ionization parameter, or
U/T) and temperature wherein a cloud of photoionized gas is stable to thermal
perturbations: at low ionization/temperature, when line cooling
is efficient, and at high temperature/ionization, when thermal balance
is achieved via Compton processes. In the intermediate region, small
changes in ionization can result in large changes in temperature. Although
the exact relationship between U/T and T for a model nebula depends on the 
SED of the 
continuum radiation, including the non-ionizing continuum which is
critical in the Compton regime, and the atomic parameters, such 
as collision strengths, instability typically occurs between 
5 x 10$^{4}$ K and 10$^{7}$ K. Both our low- and high-state models for
the core components predict that these absorbers remain thermally
stable, which may help explain their continuing presence. On the other
hand, the blueshifted components may not be stable. In Figure 7, we
show the instability curve for the screened SED used for the models
of the blueshifted components. In the current low-state, component 5 is
in the low-temperature stable regime, while the more highly ionized components
lie along the vertical, quasi-stable, region (unstable regions are places where
$dT/dU < 0$). This implies that
these components have experienced large temperature fluctuations in response
to the changes in ionizing flux, which may explain part of the discrepant
model predictions. Perhaps most of this gas is in the Compton-cooled state, 
out of which UV absorbers condense when the continuum flux drops
sufficiently. The variable opacity of the intervening core components
amplifies the changes in ionizing flux and, hence, thermal perturbations
of the blueshifted components.

\section{Summary}

We have used medium-resolution echelle spectra obtained with {\it HST}/STIS to 
study the physical conditions in the UV absorbers in the Seyfert 1 galaxy, 
NGC 3516 while it was in a recent low-continuum state. We have used 
contemporaneous {\it CXO}/LETG data (discussed in Paper I) to examine the
connection between the UV and X-ray absorbers, and have compared our
model results to archival, high-state GHRS and {\it ASCA} spectra.

 We find that the bulk of the UV and X-ray absorption arises in the same gas, principally
components 1,2, and 3$+$4. The line-of-sight covering factors for these
components are all $>$ 0.9, hence they must be screening one another 
from the ionizing source. We have demonstrated that the historical
changes in the properties of the absorbers are due to changes in their 
ionization state in response to different levels of the ionizing flux. The
changes are more dramatic for the screened components, which are
also affected by the changes in opacity of the intervening gas. In particular,
the disappearance of components 5 -- 8 in C~IV (the variable blueshifted 
absorption seen in {\it IUE} spectra) is due to the much higher ionization
parameters at high states, as a result of the much lower opacities of
components 1 --4.

  Although there are apparent similarities between NGC 3516 and NGC 4151, 
such as linear radio structure (Nagar \& Wilson 1998), NLR kinematics (Ruiz et al.
in preparation), and central source luminosity, there are striking 
differences in the appearance of their UV spectra. First, unlike NGC 4151
(see Kraemer et al. 2001a), there is little evidence for scattered
continuum radiation in a 0\arcsecpoint2 X 0\arcsecpoint2 aperture centered
on the nucleus; alternatively, the absorbers are {\it covering}
the scattering region. Second, in Kraemer et al. (2001a) we argued that the
X-ray absorption in NGC 4151 was due to a component of highly ionized gas 
which we
were unable to identify among the UV absorbers, which is clearly not the
case for NGC 3516. Finally, the absorbers in NGC 3516 are both in a higher
ionization state and of lower density than those in NGC 4151. It will be
interesting if these differences are intrinsic to these otherwise
similar AGN, or are a clue to the nature of mass outflow.

\acknowledgments

 S.B.K. and D.M.C. acknowledge support from NASA grant NAG5-4103. 
 H.N. acknowledges support by ISF and by UMBC. 
 T.J.T. acknowledges support for NASA grant NAG5-7538.
 S.B.K. thanks Gary Ferland and Fred Bruhweiler for illuminating 
 discussions. We thank Jay Dunn for help with the tables.
 We thank an anonymous referee for their careful reading of the manuscript
 and useful comments.

\clearpage

\figcaption[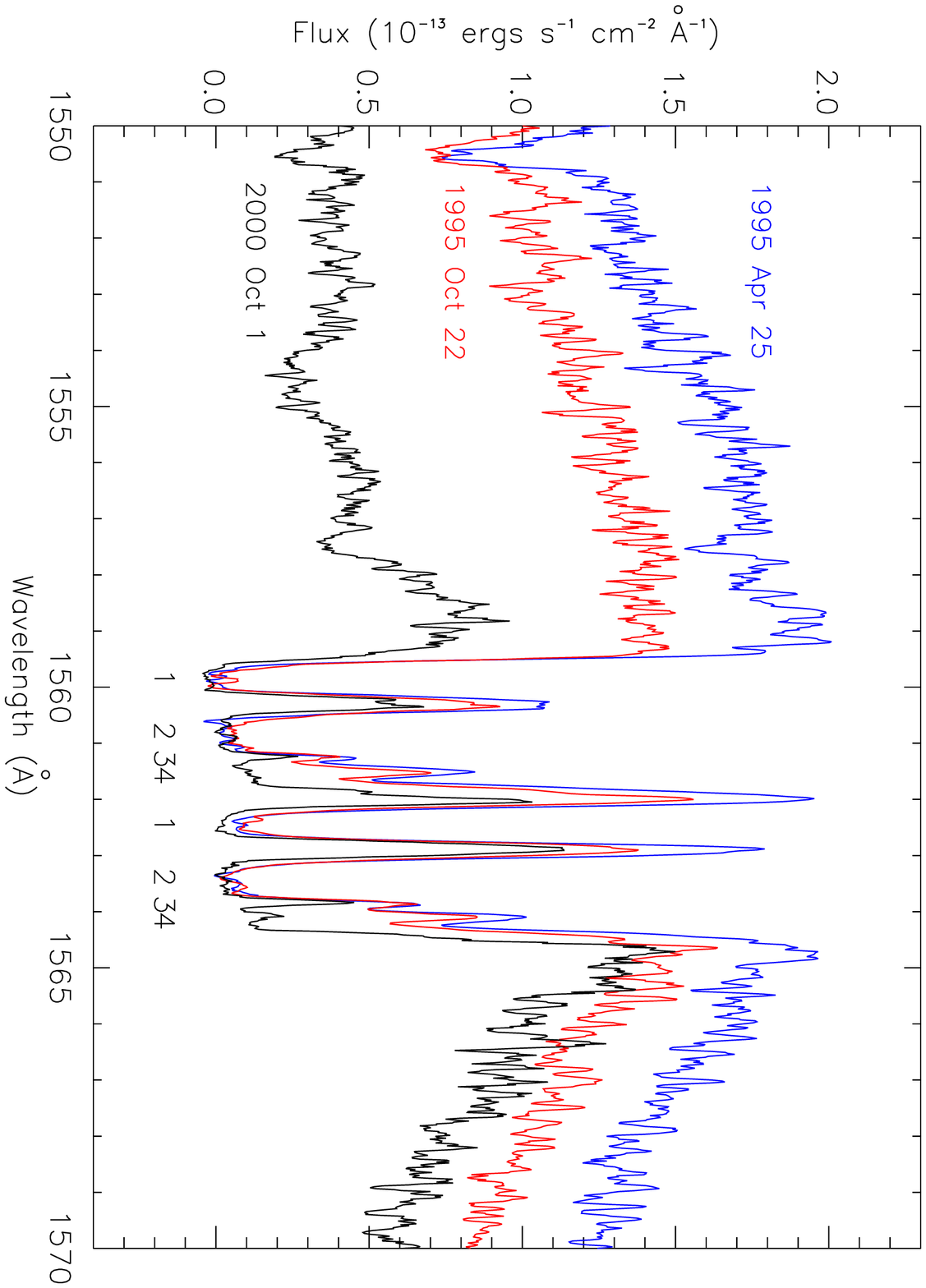]{GHRS (blue and red lines) and STIS (black line) spectra of 
the C~IV region in NGC~3516. The kinematic components of the 
intrinsic C~IV absorption doublet that were originally detected in the GHRS 
spectra are numbered.}

\figcaption[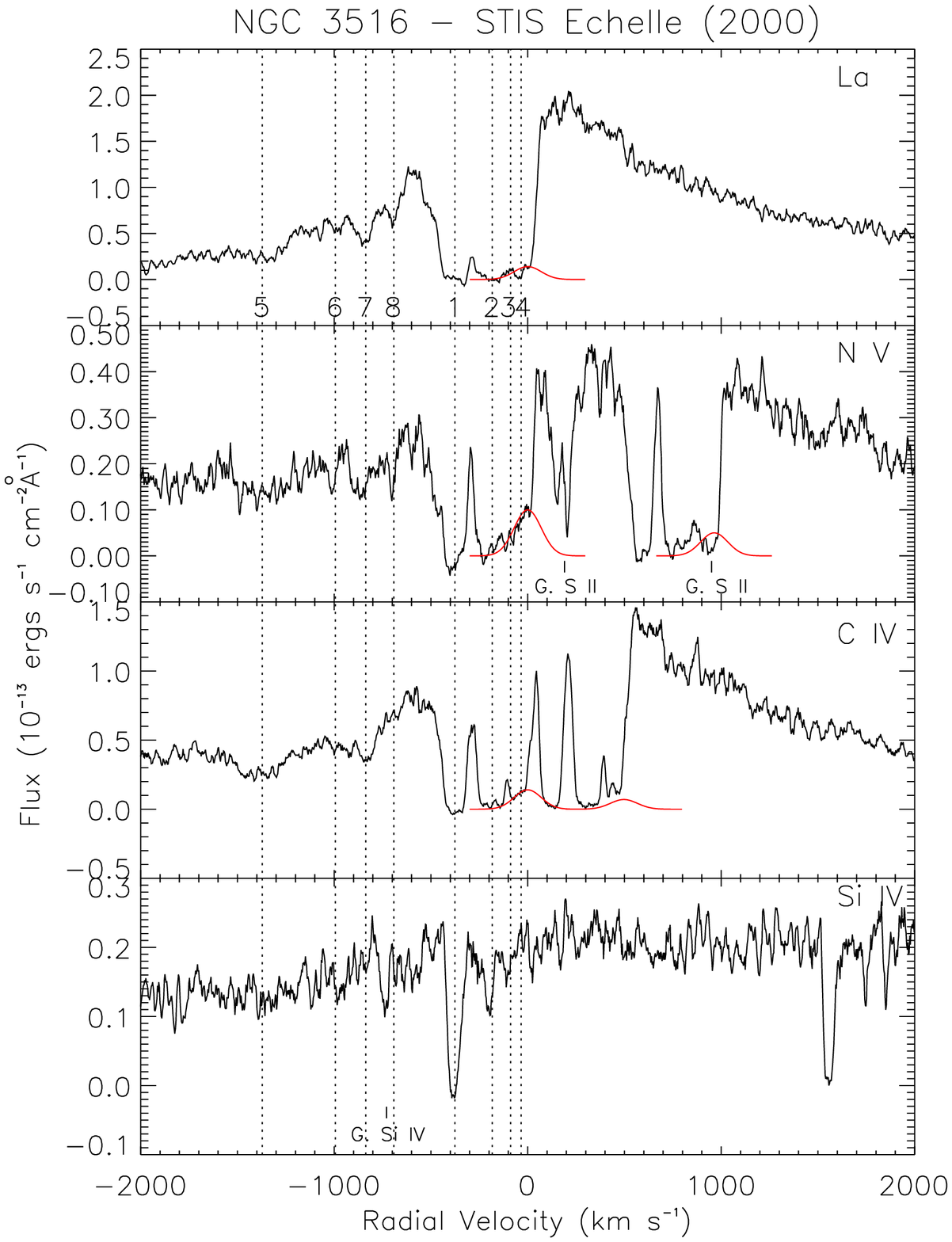]{ Portions of the STIS echelle spectra of NGC~3516, showing 
the intrinsic absorption lines in different ions. Fluxes are plotted as a 
function of the radial velocity (of the strongest member, for the doublets), 
relative to an emission-line redshift of z $=$ 0.00875. The kinematic 
components are identified for the strong members of the doublets, and vertical 
dotted lines are plotted at their approximate positions. Strong Galactic 
absorption lines are labeled.
The unocculted NLR profiles discussed in Section 2.2 are overplotted in red.
}

\figcaption[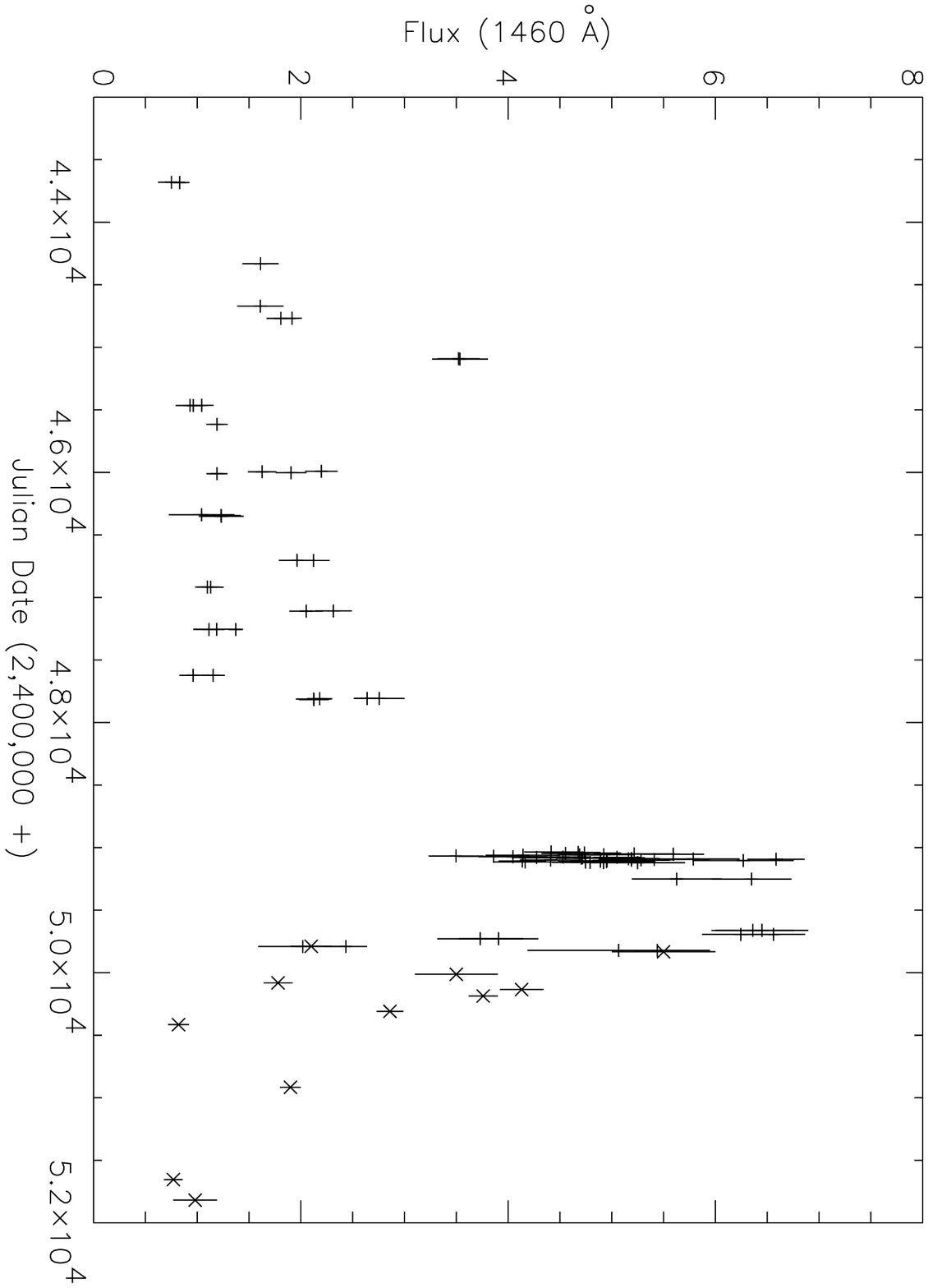]{Far-UV continuum light curve of NGC~3516. Fluxes (ergs 
s$^{-1}$ cm$^{-2}$ \AA$^{-1}$) at 1460~\AA\  are plotted as a function of 
Julian date. The pluses are from {\it IUE}, and the X's are from the other 
satellites; vertical lines indicate the error bars ($\pm$ one sigma).}

\figcaption[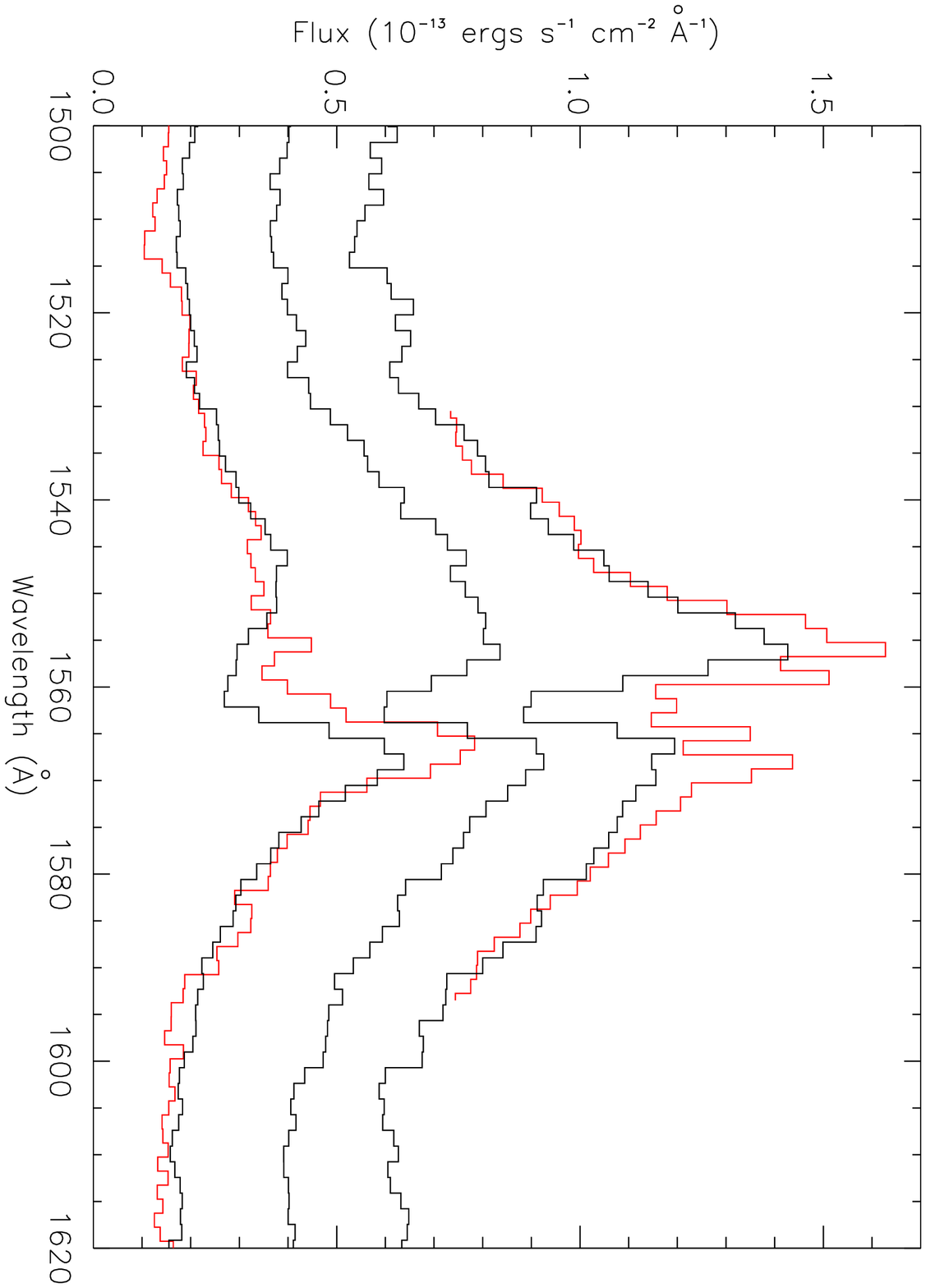]{Averaged {\it IUE} spectra in the C~IV region at three 
different continuum flux states (black lines). For comparison, rebinned 
spectra from the GHRS high state and STIS low state (red lines) 
are overplotted.}

\figcaption[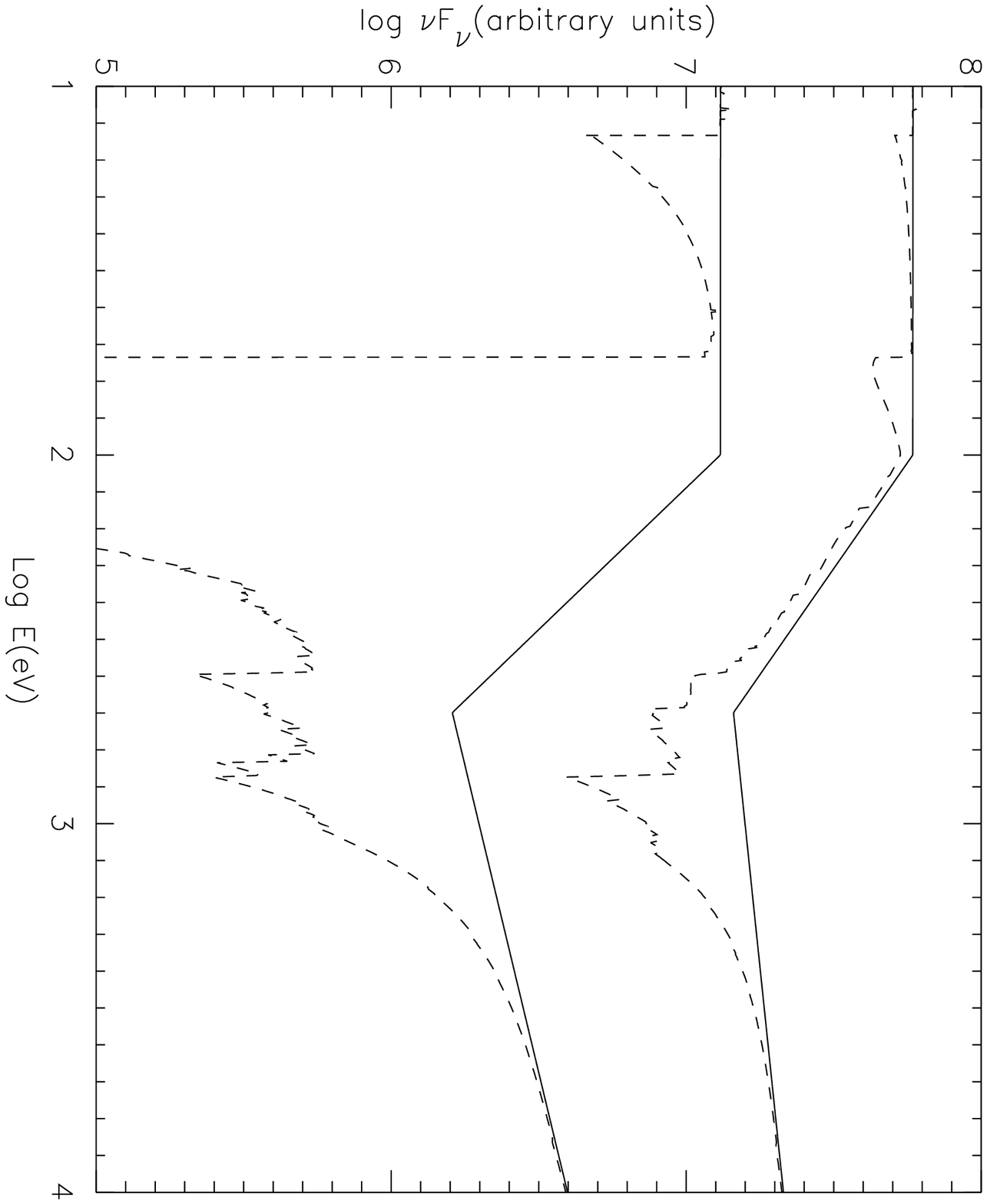]{Low (lower part of panel) and High (upper part of panel) 
state model ionizing continua. The solid 
line represents the continuum incident upon component 3$+$4, while the
dotted line is the continuum transmitted by component 1.}

\figcaption[fig6.ps]{Fits to the 1994 {\it ASCA} (high level)
and 2000 {\it CXO}/LETGS (low level) observations, based on the 
predictions of the UV absorbers models. No attempt has been made to fit the
the Fe K$\alpha$ emission near 6.4 keV.}

\figcaption[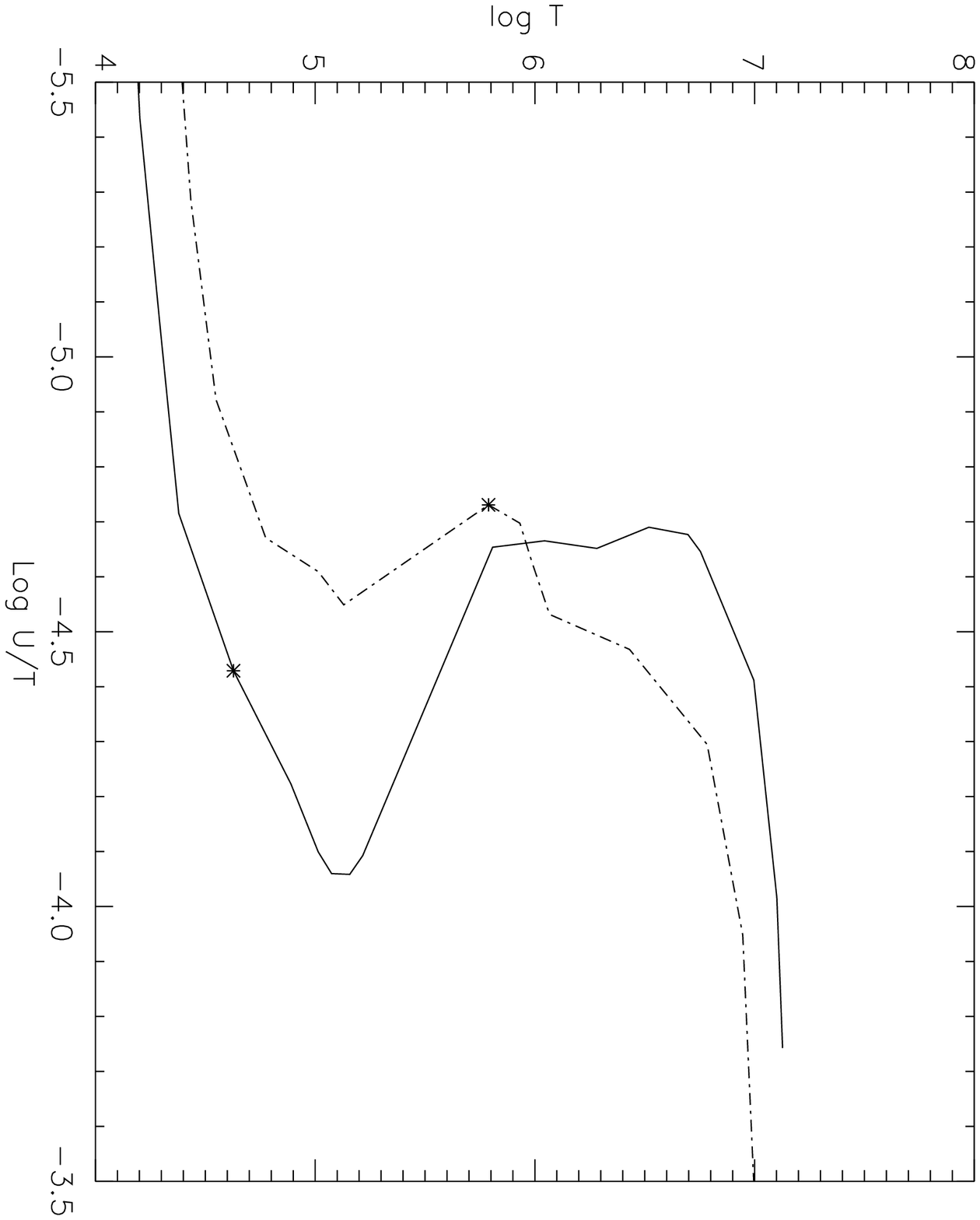]{Thermal stability curve for models generated with the
same SED used for the low-state (solid line) and high-state (dash-dotted
line) component 5 models. The flat sections, at low
and high temperature, are the stable line-cooled and Compton-cooled regions,
respectively. The symbols are the location of our best-fit models
for component 5 at the low and high flux states.
Note that the very steep rise for T is $>$ 10$^{5}$ K in the low-state
curve is due to the effects of the screened SED.}

\begin{deluxetable}{cccclll}
\tablecolumns{7}
\footnotesize
\tablecaption{{\it HST} High-Resolution Spectra of NGC 3516}
\tablewidth{0pt}
\tablehead{
\colhead{Instrument} & \colhead{Grating} & \colhead{Coverage} &
\colhead{Resolution} & \colhead{Exposure} & \colhead{Date} & \colhead{}\\
\colhead{} & \colhead{} & \colhead{(\AA)} &
\colhead{($\lambda$/$\Delta\lambda$)} & \colhead{(sec)} & \colhead{(UT)}
& \colhead{(MJD)}
}
\startdata
STIS &E140M &1150 -- 1730 &46,000 &5488   &2000 October 1 & 2451819.5 \\
STIS &E230M &2275 -- 3120 &30,000 &2220   &2000 October 1 2451819.5 \\
& & & & \\
GHRS &G160M &1529 -- 1597$^{a}$ &20,000 &7296$^{b}$    &1995 April 25 & 2449833.5\\
GHRS &G160M &1529 -- 1597$^{a}$ &20,000 &6084$^{b}$    &1995 October 22 &
2449833.5\\

\tablenotetext{a}{Covers the C~IV region (Crenshaw et al. 1998, 1999).}
\tablenotetext{b}{per grating setting; two settings were used.}
\enddata
\end{deluxetable}

\begin{deluxetable}{crcr}
\tablecolumns{4}
\footnotesize
\tablecaption{Absorption Components in NGC~3516}
\tablewidth{0pt}
\tablehead{
\colhead{Component} & \colhead{Velocity$^{a}$} & \colhead{FWHM} & 
\colhead{C$_{los}$$^{b}$}\\
\colhead{} &\colhead{(km s$^{-1}$)} &\colhead{(km s$^{-1}$)} & \colhead{}
}
\startdata
1     &$-$376 ($\pm$7)  & 70 ($\pm$8) & 0.98 ($\pm$0.02) \\
2     &$-$183 ($\pm$8)  & 44 ($\pm$10) & 0.98 ($\pm$0.02) \\
3$+$4 &$-$36  ($\pm$4)  & 20 ($\pm$4), 31 ($\pm$5)$^{c}$ & 0.95 ($\pm$0.02) \\
5     &$-$1372 ($\pm$9) & 271 ($\pm$44) & 0.81 ($\pm$0.21) \\
6     &$-$994 ($\pm$16) & 36~ ($\pm$6) & --------------- \\
7     &$-$837 ($\pm$7)  & 99 ($\pm$16) & --------------- \\
8     &$-$692 ($\pm$4)  & 35~ ($\pm$5) & --------------- \\
\tablenotetext{a}{Velocity centroid for a systemic redshift of z $=$ 0.00875.}
\tablenotetext{b}{Covering factor in the line of sight from the core of the 
C~IV doublet.}
\tablenotetext{c}{From GHRS components 3 and 4, respectively (see Crenshaw et 
al. 1999).}
\enddata
\end{deluxetable}

\begin{deluxetable}{cccccccc}
\tablecolumns{8}
\footnotesize
\tablecaption{Measured Ionic Column Densities (10$^{14}$ cm$^{-2}$)}
\tablewidth{0pt}
\tablehead{
\colhead{Comp.} & \colhead{Ly$\alpha$} & \colhead{N~V} & \colhead{C~IV} & 
\colhead{Si~IV} & \colhead{C~II} & \colhead{C~II*} & \colhead{Mg~II}
}
\startdata
 & & & STIS & & & & \\
\hline
1 & $>$5.99 & $>$20.48 & $>$10.26 & $>$1.86 & 0.79 $\pm$0.25 & 1.44 $\pm$0.30 
& 0.05 $\pm$0.02 \\
2 & $>$6.56  & $>$21.55 & $>$13.75 &0.15 $\pm$0.05 & $<$0.25 & $<$0.25 
&$<$0.02 \\
3$+$4 &$>$3.34 &$>$4.84 &$>$6.20
&$<$0.04 & $<$0.25 & $<$0.25 &$<$0.02 \\
5 &0.97 $\pm$0.18 & 1.65 $\pm$0.39 & 2.11 $\pm$0.30
&$<$0.04 & $<$0.25 & $<$0.25 &$<$0.02 \\
6 &0.09 $\pm$0.04 &0.29 $\pm$0.10 &0.10 $\pm$0.04
&$<$0.04 & $<$0.25 & $<$0.25 &$<$0.02 \\
7 &0.31 $\pm$0.05 &1.35 $\pm$0.24 &0.31 $\pm$0.07
&$<$0.04 & $<$0.25 & $<$0.25 &$<$0.02 \\
8 &0.26 $\pm$0.04 & 0.48 $\pm$0.14 & $<$0.07
&$<$0.04 & $<$0.25 & $<$0.25 &$<$0.02 \\
\hline
 & & & GHRS & & & & \\
\hline
1 & & &$>$8.0 & & & & \\
2 & & &$>$13.7 & & & & \\
3 & & &0.4$ \pm$0.1 & & & & \\
4 & & &0.5$ \pm$0.1 & & & & \\
\enddata
\end{deluxetable}

\begin{deluxetable}{lcrll}
\tablecolumns{5}
\footnotesize
\tablecaption{Low-Resolution FUV Spectra of NGC 3516}
\tablewidth{0pt}
\tablehead{
\colhead{Instrument/} & \colhead{Coverage} &
\colhead{Resolution} & \colhead{Date} & \colhead{}\\
\colhead{Grating} & \colhead{(\AA)} &
\colhead{($\lambda$/$\Delta\lambda$)} & \colhead{(UT)} & \colhead{(MJD)}
}
\startdata
IUE SWP    &1150 -- 1978 &$\sim$250 &1978 June 20 -- 1995 April 14$^{a}$
& 2443679.5 -- 2449821.5 \\
HUT$^{b}$  &~820 -- 1840 &$\sim$450 &1995 March 11 -- 13 
& 2449787.5 -- 2449789.5 \\
FOS G130H & 1150 -- 1605 &$\sim$1200 &1995 December 30 & 2450081.5 \\
FOS G130H & 1150 -- 1605 &$\sim$1200 &1996 February 21 & 2450134.5 \\
FOS G130H & 1150 -- 1605 &$\sim$1200 &1996 April 13 & 2450186.5 \\
FOS G130H & 1150 -- 1605 &$\sim$1200 &1996 August 14 & 2450309.5 \\
FOS G130H & 1150 -- 1605 &$\sim$1200 &1996 November 28 & 2450415.5\\
STIS G140L & 1150 -- 1715 &$\sim$1000 &1998 April 13 & 2450916.5 \\
STIS G140L & 1150 -- 1715 &$\sim$1000 &2000 April 17 & 2451651.5 \\

\tablenotetext{a}{See the IUE Merged Log at http://archive.stsci.edu/iue.}
\tablenotetext{b}{Hopkins Ultraviolet Telescope (Kruk et al. 1995)}
\enddata
\end{deluxetable}

\begin{deluxetable}{lccc}
\tablecolumns{4}
\footnotesize
\tablecaption{Model Parameters}
\tablewidth{0pt}
\tablehead{
\colhead{Model} & \colhead{U} & \colhead{N$_{H}$ (cm$^{-2}$)} 
& \colhead{$<T_{e}>$ (K)}
}
\startdata
 & Low-state models & &    \\
\hline
1A & 0.126 & 2.4e21 & 2.30e04 \\
1B & 0.001 & 7.0e18 & 1.61e04 \\
2  & 0.138 & 3.9e21 & 2.50e04 \\
3$+$4 & 0.185 & 1.4e21 & 2.78e04 \\
5 & 1.58 & 1.5e19 & 4.25e04 \\
6 & 8.50 & 1.5e20 & 1.05e05 \\
7 & 10.7 & 2.0e21 & 1.39e05 \\
8 & 10.7 & 5.5e20 & 1.26e05 \\
\hline
 & High-state models & &   \\
\hline
1A & 0.793 & 2.4e21 & 4.74e04 \\
1B & 0.008 & 7.0e18 & 1.89e04\\
2 & 0.669 & 2.4e21 & 4.86e04\\
3$+$4 & 0.845 & 1.4e21 & 5.69e04 \\
5 & 11.43 & 1.5e19 & 6.15e05 \\
\enddata
\end{deluxetable}

\begin{deluxetable}{lcccccc}
\tablecolumns{7}
\footnotesize
\tablecaption{Predicted Ionic Column Densities$^{a}$ (10$^{14}$ cm$^{-2}$)}
\tablewidth{0pt}
\tablehead{
\colhead{Model} & \colhead{Ly$\alpha$} & \colhead{N~V} & \colhead{C~IV} & 
\colhead{Si~IV} & \colhead{C~II} & \colhead{Mg~II}
}
\startdata
 & & & Low-state models (STIS epoch) & & &  \\
\hline
1A & 7.52e02 & 6.85e01 & 2.03e03 & 1.96 & 0.09 & 0.05\\
1B & 3.06e02 & 0.03 & 2.72 & 0.46 & 2.23 & 0.14\\
   & ($>$5.99) & ($>$20.48) & ($>$10.26) & ($>$1.86) & (2.23 $\pm$0.39) 
& (0.05 $\pm$0.02) \\
2 & 1.02e03 & 7.40e02 & 5.22e02 & 0.20 & 0.01 & -- \\
  & ($>$6.56)  & ($>$21.55) & ($>$13.75) & (0.15 $\pm$0.05) & ($<$0.25) 
& ($<$0.02) \\
3$+$4 & 2.87e02 & 1.42e02 & 70.1 & -- & -- & -- \\
 & ($>$3.34) & ($>$4.84) & ($>$6.20)
& ($<$0.04) & ($<$0.25) & ($<$0.02) \\
5 & 0.30 & 1.59 & 2.13 & -- & -- & -- \\
  & (0.97 $\pm$0.18) & (1.65 $\pm$0.39) & (2.11 $\pm$0.30)
& ($<$0.04) & ($<$0.25) & ($<$0.02) \\
\hline
 & & & High-state models (GHRS epoch) & & &  \\
\hline
1A & & & 4.58 & & & \\
1B & & & 9.37 & & & \\
  & & & ($>$8.0) &  & & \\
2 & & & 8.7 &  & & \\
  & & & ($>$13.7) &  & & \\
3$+$4 & & & 1.0  & & \\
  & & & (0.9 $\pm$0.14) &  & & \\
\tablenotetext{a}{The values in parentheses are the measured column 
densities (see Table 3).}
\enddata
\end{deluxetable}

\clearpage
\vskip3.0in
\begin{figure}
\plotone{f1.eps}
\\Fig.~1.
\end{figure}

\clearpage
\vskip3.0in
\begin{figure}
\plotone{f2.eps}
\\Fig.~2.
\end{figure}

\clearpage
\vskip3.0in
\begin{figure}
\plotone{f3.eps}
\\Fig.~3.
\end{figure}

\clearpage
\vskip3.0in
\begin{figure}
\plotone{f4.eps}
\\Fig.~4.
\end{figure}

\clearpage
\vskip3.0in
\begin{figure}
\plotone{fig5.ps}
\\Fig.~5.
\end{figure}


\clearpage
\vskip3.0in
\begin{figure}
\plotone{fig7.ps}
\\Fig.~7.
\end{figure}

\end{document}